\documentstyle[sprocl,epsfig,amsmath]{article}

\def\np#1#2#3{Nucl.\ Phys.\ B#1 (19#3) #2}
\def\pl#1#2#3{Phys.\ Lett.\ #1B (19#3) #2}

\def\zp#1#2#3{Zeit.\ Phys.\ C#1 (19#3) #2}
\def\cA{{\cal{A}}}
\def\cB{{\cal{B}}}
\def\cC{{\cal{C}}}

\def\as{\alpha_{\mbox{\scriptsize s}}}
\def\asb{{\bar\alpha}_{\mbox{\scriptsize s}}}
\begin{document}
\begin{flushright}
  IFUM-622-FT \\
  hep-ph/9805322 \\
  May 1998
\end{flushright}
\vspace{1cm}

\title{Associated quantities from the CCFM approach\footnote{Talk
    presented at DIS98, 6th International Workshop on Deep Inelastic
    Scattering, Brussels, April 1998.}}

\author{G.P. Salam}

\address{INFN --- Sezione di Milano, Via Celoria, Milano 20133,
  Italy\\E-mail: Gavin.Salam@mi.infn.it}  

\maketitle\abstracts{Results are presented on structure functions and
  final state properties within the CCFM approach. Traditionally used
  forms of the CCFM equation have difficulty fitting the $F_2$ data,
  predicting too fast a growth at small $x$. A solution can be found
  in a particular treatment of formally subleading $(1-z)$ terms,
  which dampens very considerably the small-$x$ growth. Preliminary
  results are shown for the transverse energy flow, and future
  prospects and plans are discussed.}

\section{Introduction}
\label{sec:intro}
This talk presents a summary of the results obtained during the past
year in collaboration with Bottazzi, Marchesini and Scorletti on the
predictions for structure functions and final state properties from
the CCFM equation.\cite{CCFM} 

Like the BFKL equation,\cite{BFKL} the CCFM equation resums
logarithms of $x$, but in contrast, as a consequence of its inclusion
of angular ordering of initial-state radiation, it also
takes into proper account $\ln 1/x$ terms associated with the
final-state. In particular, for a number of final state properties
(e.g.\ multiplicities), the correct small-$x$ perturbative result, as
given by the CCFM equation, contains terms of the form
$$
(\as \ln^2 1/x)^n\,,
$$
whereas in calculations neglecting the angular ordering (BFKL),
one obtains terms of the form
$$
(2\as \ln 1/x \ln Q/\mu)^n\,,
$$
where $\mu$ is an infra-red cutoff. Even in final-state quantities
which are the same at leading order, angular ordering can introduce
phenomenologically very substantial next-to-leading corrections.

\section{Angular ordering}
\label{sec:ao}
\begin{figure}[htbp]
  \begin{center}
    \scalebox{0.7}{\input{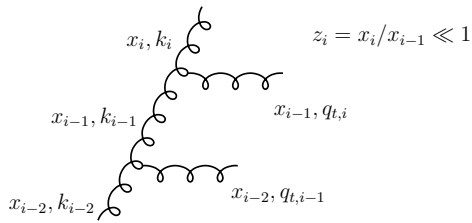}}
    \caption{Kinematics}
    \label{fig:kine}
  \end{center}
\end{figure}

It is essential for what follows to consider the kinematics of angular
ordering. In figure \ref{fig:kine}, the angle $\theta_i$ of gluon $i$
is given by the following equation
$$
p_i = x_i E_p \tan \theta_i = \frac{z_i q_{t,i}}{1-z_i},
$$
with $E_p$ the energy of the proton. For simplicity,
one works in terms of a rescaled transverse momentum $q_i =
q_{t,i}/(1-z_i)$. One then obtains the following equation for the
unintegrated gluon density $\cA(x,k,p)$, where the third variable $p$
limits the maximum angle of gluon emission during the evolution:
\begin{equation}
  \label{eq:AKA}
  \cA(x,k,p) = \cA^{(0)}(x,k,p) + K\otimes\cA,
\end{equation}
$$
K\otimes\cA \equiv \int \frac{dz}{z} \frac{d^2q}{\pi q^2} \asb[(1-z)q]\;
\Theta(p-zq)\;\Delta(z,q,k)\;\cA(x/z,|k+(1-z)q|,q) \,.
$$
Here, $\cA^{(0)}$ is the initial condition, $\asb =  \as N_C/\pi$,
and the form factor $\Delta$, which resums virtual corrections, is
\begin{equation}
  \label{eq:ff}
  \ln \Delta(z,q,k)=
  -\int_{z}^1 \frac{dz'}{z'} \int\frac{d^2{q'}}{\pi{q'}^2}\asb[(1-z')q']
  \;\Theta(k-Tq')\;\Theta(q'-z'q)
  \,.
\end{equation}
Traditionally, the factor $T$ is taken as $T=1$, but given that $q'$
is in reality a scaled transverse momentum, $q' = q'_t/(1-z')$, it is
equally reasonable to have $T=(1-z)$, in analogy with the $|k+(1-z)q|$
in $K\otimes\cA$. Most previous calculations have actually
ignored all $(1-z)$ factors, making the approximation $(1-z)\to1$ on
the grounds that $z\ll1$, and that the resulting effect is at
most NLL. This was the approach initially adopted also by our group.

\section{Structure functions}
\label{sec:sf}
\begin{figure}[tbp]
  \begin{center}
    \scalebox{0.8}{\input{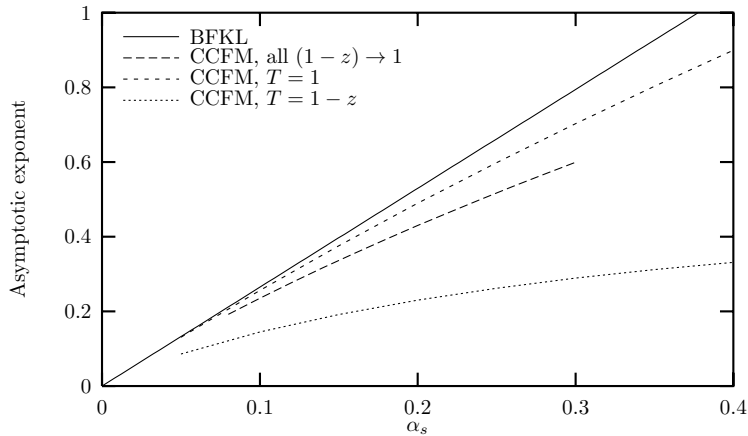}}
    \caption{The asymptotic exponents, $-d \ln F_2/d \ln x$, for fixed
      $\as$, as a function of $\as$; shown for various small-$x$
      evolution equations.} 
    \label{fig:exp}
  \end{center}
\end{figure}

As a first step, and as a check of the consistency of the whole
procedure, we fitted the HERA data\cite{HERAF2} for the structure
function $F_2(x,Q^2)$ in the region $x<10^{-2}$, $8<Q^2<150$~GeV$^2$.
The parameters which were included in the fit were the initial
condition, the lowest allowed transverse momentum and the value of
$\as$ at which it ``freezes''. Typical best fits had a
$\chi^2/$d.o.f.$\simeq 10$, mainly because $F_2$ rises far more slowly
than is given by the CCFM equation: $F_2$ rises with an exponent
$0.2$--$0.3$, while the exponent from the CCFM equation ($(1-z)\to1$
approximation), plotted against $\as$ in figure~\ref{fig:exp}, for the
relevant range of $\as$ is simply too high.

At this point one is induced to examine the effect of treating $(1-z)$
properly. Figure~\ref{fig:exp} shows the two possibilities, according
to one's choice for $T$ in \eqref{eq:ff}. For $T=1$, there is
relatively little change, while for $T=(1-z)$ the exponent is
drastically reduced, and one quite easily obtains a good fit
($\chi^2/$d.o.f.$\simeq1$) to the $F_2$ data.

The large effect of such a formally NLL term is not all that
surprising given one's knowledge of the magnitude of the full NLL
kernel.\cite{NLL} Nevertheless it translates into a significant
uncertainty on any prediction from the CCFM approach, at least until
one is able to understand the full NLL kernel in the context of the
CCFM equation. The way in which we have decided to go forwards, in the
face of such uncertainty, is to choose the form of the equation
($T=1-z$) which allows one to fit the structure function, and from
there to go and examine final state properties, also known as
associated quantities.

\section{Associated quantities}
\label{sec:aq}

The method of associated quantities allows one to determine final
state properties through the following steps. One determines the
unintegrated gluon density (for all relevant $x$, $k$, $p$) as usual
by solving \eqref{eq:AKA}. One then acts on it with a ``reduced''
kernel $K_D$ which corresponds to allowing one emission which goes
into a detector $D$:
$$
\cB = K_D \otimes \cA.
$$
Finally one obtains a gluon density $\cC$ which includes
any number of further emissions by solving the integral 
equation (analogous to \eqref{eq:AKA})
$$
\cC = \cB + K \otimes \cC.
$$
\begin{figure}[tbp]
  \begin{center}
    \scalebox{0.8}{\input{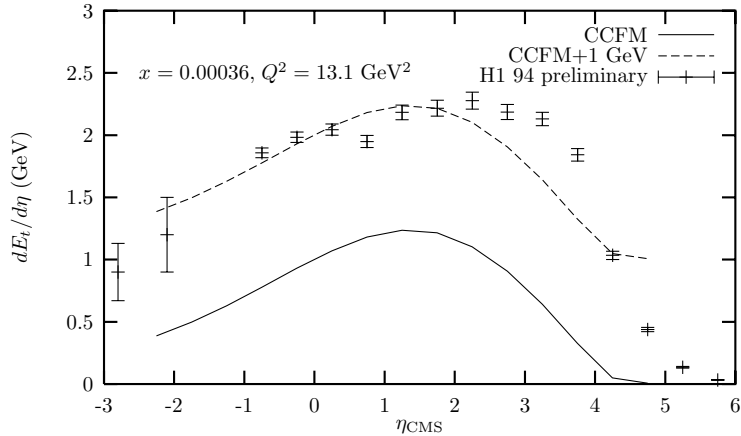}}
    \caption[]{$E_t$ flow as a function pseudo-rapidity in the hadronic
      centre-of-mass system, neglecting the contribution from the
      $q{\bar q}$ box, compared with H1 data.\cite{Kuhlen}}
    \label{fig:Et}
  \end{center}
\end{figure}
Preliminary results are just becoming available, and the $E_t$ flow is
shown in figure~\ref{fig:Et}. The agreement with the data is rather
poor; possible reasons are that hadronisation corrections and initial
state radiation of soft gluons, both of which may contribute
significantly, are not taken into account. Adding uniformally $1$~GeV
of radiation to simulate hadronisation effects leads to reasonable
agreement, but this amount is somewhat large for comfort, and in any
case one should really test such a procedure at other $x$ and $Q^2$
values as well. In the near future we expect to calculate other final
state properties, such as the forward-jet cross section and the
$k_t$-spectrum of charged particles, both of which are expected to be
somewhat less sensitive to hadronisation and initial-state soft gluon
effects. 

\section{Conclusions and outlook}
For the CCFM equation, formally subleading $(1-z)$ terms have a very
large effect on structure function predictions. Choosing them so as to
reproduce $F_2$, allows one to go on and examine final state
properties; some results ($E_t$ flow) are already available, more will
come in the near future.\cite{BMSS} It should be borne in mind that
the particular NLL choice that we have made is quite arbitrary, and
that other NLL choices might equally well reproduce $F_2$, but give
different final-state properties --- to do any better one needs to
know how to incorporate the full NLL kernel~\cite{NLL} into the CCFM
equation, a non-trivial operation.

Is there any point in doing phenomenology without including the full
NLL kernel? The answer is perhaps ``yes'': if a first NLL effect kills
most of the initial-state radiation, then a second one, which on its
own might have been very important, will have little radiation left to
kill, and so have little impact. This leaves the hope that even if
only some of the NLL effects are included, one may still have a
reasonable description of small-$x$ physics.

\section*{Acknowledgements}
This work was carried out in collaboration with G. Bottazzi, G.
Marchesini and M. Scorletti. We are grateful to J. Bartels, T. Carli,
M.  Ciafaloni, Yu.L.  Dokshitzer, M. Kuhlen and B.R. Webber for
helpful discussions.

\end{document}